\documentclass[twocolumn,preprintnumbers,amsmath,superscriptaddress,amssymb]{revtex4}

\usepackage{graphicx}
\usepackage{dcolumn}
\usepackage{bm}

\begin{document}

\title{Nonuniform growth and topological defects in the shaping of elastic sheets}
\author{Nakul P. Bende}
\author{Ryan C. Hayward}
\affiliation{Department of Polymer Science, University of Massachusetts, Amherst MA, 01003}
\author{Christian D. Santangelo}
\affiliation{Department of Physics, University of Massachusetts, Amherst MA, 01003}
\date{\today}

\begin{abstract}
Programming the non-uniform growth of a responsive polymer gel has emerged as a powerful tool to shape sheets into prescribed three dimensional shapes. We demonstrate that shapes with zero Gaussian curvature, except at singularities, produced by the growth-induced buckling of a thin elastic sheet are the same as those produced by the Volterra construction of topological defects in which edges of an intrinsically flat surface are identified. With this connection, we study the problem of choosing an optimal pattern of growth for a prescribed developable surface, finding a fundamental trade-off between optimal design and the accuracy of the resulting shape which can be quantified by the length along which an edge should be identified.
\end{abstract}

\maketitle

Non-uniform growth processes in elastic sheets have been exploited to create a wide variety of target shapes \cite{sharon, hayward}. The underlying concept is that spatially nonuniform growth induces in-plane stresses which are relieved if the sheet buckles. In the limit of infinitesimal thickness, the resulting Gaussian curvature is determined entirely by the pattern of growth \cite{lewicka}. Yet we may also produce Gaussian curvature by removing a wedge from a sheet of paper and identifying the newly cut edges. the paper will then buckle into a cone with singular Gaussian curvature at the tip \cite{mmm,dipole}. This illustrates the deep relationship between disclinations and Gaussian curvature \cite{seung, kupf}, and has wide-ranging implications in the faceting of viruses \cite{virus} and fullerenes \cite{fullerene}, shape transitions in protein-coated cell membranes \cite{clathrin}, and buckling in graphene \cite{graphene}. Indeed, the engineering of crystalline defects has been proposed as a means to control shape elastically \cite{nelson}.

In this letter, we show that these two superficially different processes produce identical shapes in thin, elastic films for surfaces with point singularities of Gaussian curvature. Thus, non-uniform but isotropic growth is equivalent, in a precise sense, to the Volterra construction of disclinations and dislocations \cite{volterra}. Indeed, in the limit of small curvatures, generalizations of the F\"oppl-von K\`arm\`an equations show that defects \cite{seung} and nonuniform growth \cite{liang} enter the equilibrium equations in the same manner, implying that they produce the same shape.

Making this relationship explicit, as we have done, has far reaching consequences: not only can we exploit the mapping to predict shapes resulting from nonuniform growth using only ``paper,'' scissors and adhesive, inverting the mapping allows us to find growth patterns that can be more optimally implemented within a finite-range of growth. Yet, we identify an antagonistic trade-off between the range of growth and the resolution of the growth pattern. This provides us with limits to what can be realistically achieved experimentally with non-uniform growth.
 
\begin{figure}[b]
\includegraphics[width=3.5in]{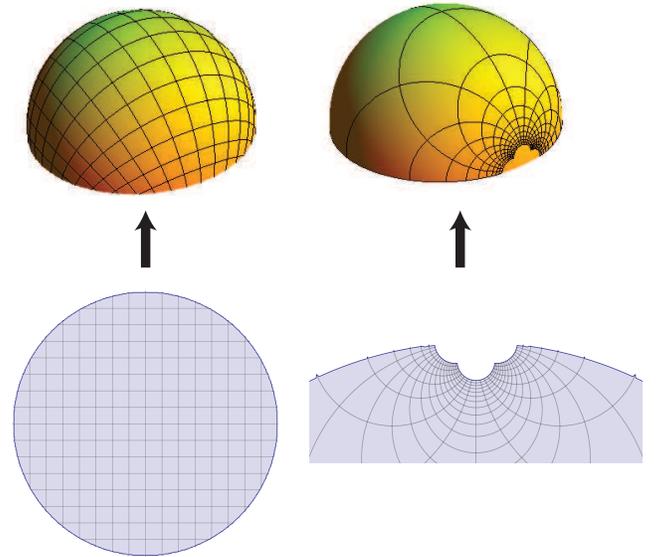}
\caption{\label{fig:map} (color online) Two domains, both of which can be mapped to the same sphere with different growth patterns. The mapping between the two is $w=g(z) = z/(1-z)$.}
\end{figure}
Isotropic growth can be described in terms of a prescribed metric, given by $ds^2 = \Omega(x,y) (dx^2+dy^2)$, where $\Omega(x,y)$ gives a multiplicative increase in area \cite{marder}. The function $\Omega(x,y)$ can be controlled, for example, through the local monomer \cite{sharon} or cross link density \cite{hayward} of a polymer gel that, subsequently, swells in a solvent. The resulting equilibrium shape is then described by a solution to a set of covariant equations (see, for example, \cite{muller,efrati,dias}), the precise form of which is not needed in our analysis here.
Covariance of the equilibrium equations imply the simple result that the solutions are invariant with respect to changing how points on the surface are labeled. Therefore, if we introduce a new coordinate system $(u,v)$ such that the growth is described by the metric $ds^2 = \tilde{\Omega}(u,v) [du^2 + dv^2]$, the physical shape of the solution remains invariant. Starting from the same buckled surface, then, we can produce two different flat domains by ``ungrowing'' according to either $1/\Omega(x,y)$ or $1/\tilde{\Omega}(u,v)$ (Fig. \ref{fig:map}). Thus, we arrive at our first mathematical result: the $(u,v)$ domain and growth pattern $\tilde{\Omega}(u,v)$ and the $(x,y)$ domain with growth pattern $\Omega(x,y)$ yield the same three-dimensional solution to the equilibrium equations (Fig. \ref{fig:map}).

For isotropic growth, the mapping between $(x,y)$ and $(u,v)$ is most compactly expressed using complex coordinates $z=x+i y$ and $w=u+i v$. In a domain of the complex $w-$plane, $ds^2 = 2 \Omega(w,\overline{w}) dw d\overline{w}$. Similarly, we define an analytic function $g(z)$ such that $w=g(z)$ maps a domain in the $z-$plane to one in the $w-$plane. It follows that the prescribed metric is $ds^2 = 2 \Omega[g(z),\overline{g(z)}] |\partial g(z)|^2 dz d\overline{z}$, where $\partial \equiv (\partial_x - i \partial_y)/2$. Consequently, a domain in the $z-$plane with isotropic growth
\begin{equation}
\tilde{\Omega}(z,\overline{z}) = \Omega[g(z),\overline{g(z)}] |\partial g(z)|^2
\end{equation}
will produce the same shape as the domain $g(z)$ in the $w-$plane with isotropic growth $\Omega[w,\overline{w}]$.

In what follows, let us restrict ourselves to isotropic growth processes resulting in surfaces with zero Gaussian curvature, $K$, except at isolated singularities. According to Gauss' \textit{theorema egregium}, one has
\begin{equation}\label{eq:Omega1}
\nabla^2 \ln \Omega = -2 K \Omega = - 2 \sum_i K_i \delta^2(\mathbf{x}-\mathbf{x}_i) \Omega(\mathbf{x}_i),
\end{equation}
so that the total Gaussian curvature of the surface is $\int dA~ K = \int d^2x~\Omega(\mathbf{x}_i) K = \sum_i K_i$. Thus, we obtain the metric
\begin{equation}\label{eq:Omega}
\Omega(z,\overline{z}) = |e^{h(z)}| \prod_i \left| z-z_i \right|^{-K_i/\pi},
\end{equation}
where $h(z)$ is an arbitrary, analytic function.

To proceed, start with a domain in the $w-$plane having metric, $ds^2 = 2 dw d\overline{w} = 2 |\partial g(z)|^2 dz d\overline{z}$. Thus,
\begin{equation}\label{eq:mapping}
\partial g(z) = e^{h(z)/2} \prod_i (z-z_i)^{-K_i/(2 \pi)},
\end{equation}
defines a mapping $g(z)$ from a domain in the $w-$plane with $\Omega(w,\overline{w}) = 1$ to a domain in the $z-$plane with metric given by Eq. (\ref{eq:Omega}). This result, in combinations with the general covariance of the elastic equations implies that a pattern of growth, corresponding to a surface with $K=0$ except at distinct singularities, buckles into the same shape as a domain with no growth at all but with some sides identified.

\begin{figure}[t]
\includegraphics[width=3.5in]{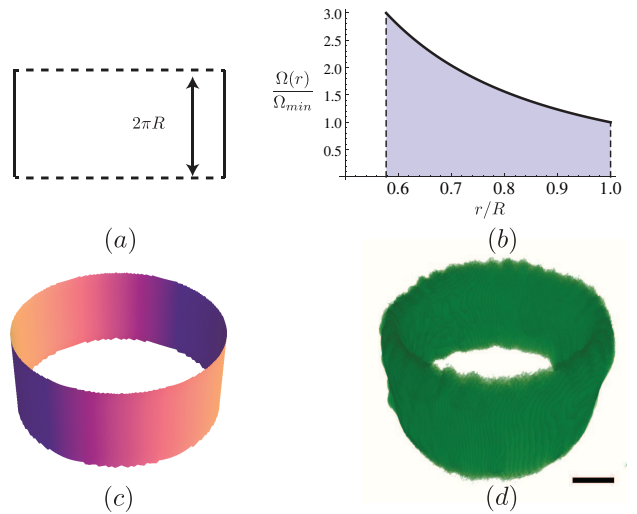}
\caption{\label{fig:cyl} (color online) (a) A domain with two opposite sides identified becomes a cylinder. (b) An annular domain with metric $\Omega = (R/r)^2$ also buckles into a cylinder, as seen by (c) numerical minimization of a bead and spring model in and (d) 3D reconstructed images from an experiment using halftone gel lithography (scale bars: 200 $\mu$m) (Supplementary Information M1).}
\end{figure}

The simplest example with which to illustrate the equivalence is the growth pattern, $\Omega(z,\overline{z}) = R^2/|z|^2$, defined on an annulus of inner radius $r_0$. The mapping to the $w-$plane is given by (Fig. \ref{fig:cyl}),
\begin{equation}
g(z) = \int dz~\frac{R}{z} = R \ln (z/R).
\end{equation}
Since the logarithm is not single-valued, one must define a branch cut on the complex plane across which $g(z)$ will be discontinuous. One might choose, for example, a branch cut along the positive real axis. Setting $z=r e^{i \theta}$ with $0 \le \theta < 2 \pi$ then implies
\begin{equation}\label{eq:equiv}
g(r e^{i \theta}) = R \ln (r/R) + i R \theta.
\end{equation}
Therefore, the region just above the positive real axis is mapped to the real axis in the $w-$plane while the line just below the positive real axis is mapped to the line $x + i 2 \pi R$. Since these two regions are connected across the positive real axis in the $z-$plane, the mapping requires us to identify the real axis with the line at $2 \pi i R$ in the $w-$plane. This, of course, is a standard construction for a cylinder.

To confirm this, we performed simulations of a film described by a system of points connected by springs. The growth pattern is encoded by choosing equilibrium spring lengths according to $\Omega(z,\overline{z})$, as described in Refs. \cite{marder,chen}. The resulting shape is, indeed, cylindrical and, even though it was produced from an annulus with a very different growth pattern near the inner and outer boundary, the result is nevertheless reflection symmetric (Fig. \ref{fig:cyl}c). Even thick cylinders, which show a gently flaring at both edges, maintain the symmetry one would expect from the mathematical equivalence encoded in Eq. (\ref{eq:equiv}). Fig. \ref{fig:cyl}d shows a cylindrical film produced from an initially flat annulus, using halftone gel lithography to pattern the swelling of a photo-crosslinkable polymer film of poly(N-isopropylacrylamide), as described in \cite{hayward}.

Eq. (\ref{eq:mapping}) provides us with a connection between Gaussian curvature and the Volterra construction of a disclination formed by removing a wedge of fixed angle. To make this connection explicit, consider an annulus with $\Omega = |z/R|^{-K/\pi}$, which buckles into a cone (see, for example, Ref. \cite{hayward}). This shape is equivalent to a domain with no growth under the mapping
\begin{equation}
g(z) = \frac{(z/R)^{1-K/(2 \pi)}}{1-K/(2 \pi)}.
\end{equation}
Again, there is a branch point at the origin and infinity; choosing the branch cut along the positive real axis, we find that we must identify the two radial lines across a wedge of angle $K$.
More generally, we see that singularities of Gaussian curvature can naturally be identified with a Schwarz-Christoffel-like transformation 
\begin{equation}\label{eq:map}
g(z) = \int dz~\prod_i (z-z_i)^{-(\bar{\theta}_i-\theta_i)/(\pi + \bar{\theta}_i)}
\end{equation}
where a wedge of angle $\theta$ becomes one of angle $\bar{\theta}$. When $\bar{\theta}=\pi$ we obtain the traditional Schwarz-Christoffel transformation; in our correspondence, however, we require $\bar{\theta}=0$.

\begin{figure}
\includegraphics[width=3.3in]{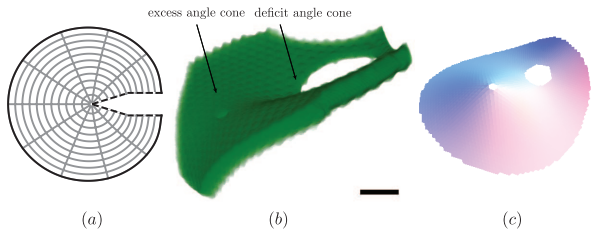}
\caption{\label{fig:dipole} (color online) (a) Domain in the $w-$plane for a dipole for $K |z_0|^2/(2 \pi) = -10^{-1}$. Also shown is the dipole formed by the corresponding growth pattern using (b) halftone gel lithography (scale bar: 200 $\mu$m) (Supplementary Information M2) and (c) numerical minimization. Only one of the two possible dipolar geometries is seen \cite{dipole}.}
\end{figure}
A pair of oppositely-charged singularities, corresponding to two opposite disclinations, can be formed from the growth pattern $\Omega(z,\overline{z}) = |z-z_0|^{-K/\pi} |z+z_0|^{K/\pi}$. The mapping $\partial g(z) = (z-z_0)^{-K/(2 \pi)} (z+z_0)^{K/(2\pi)}$ has three branch points: $z_0$, $-z_0$, and $\infty$. Choosing branch cuts from $z_0$ to infinity and $-z_0$ to infinity, we come to the intuitive conclusion that the resulting shape requires us to remove a wedge of angle $K$ from one singularity and add a wedge of angle $K$ to the other. The shapes of such structures have been studied in Ref. \cite{dipole}.

Yet, in a 2D crystal a pair of opposite disclinations corresponds to a dislocation in which a row of atoms is removed. To see that this is also the case here, consider choosing a single branch cut from $-z_0$, passing through $z_0$ and continuing to infinity. This shape requires only one cut which, as seen in Fig. \ref{fig:dipole}(a) requires only the removal of material. Indeed, far from the singularities, the resulting branch cut appears to be equivalent to the removal of a constant width of material, exactly the way an edge dislocation requires the removal of a row of atoms from a 2D crystal. To see this more clearly, we can study the far-field limit by taking $z_0 \rightarrow 0$ while $K |z_0|$ simultaneously remains constant. Then $g(z) =z - (z_0 K/\pi) \ln (z/|z_0|)$, and the Burger's vector of the edge dislocation corresponding to Fig. \ref{fig:dipole} is $i 2 K |z_0|/\pi$. Continuing in this manner, a pair of dislocations can be formed with only an internal branch cut, which forces us to remove material from the interior of a domain.

Finally, we discuss how to produce the inverse mapping from a flat domain to a growth pattern. As a concrete example, we consider forming a tetrahedron by folding an equilateral triangle. The growth pattern is described by
\begin{equation}\label{eq:omegaftet}
\Omega(z,\overline{z}) = \Omega_0 |e^{h(z)}|^2 |z^3-R^3|^{-1}.
\end{equation}
As there are many growth patterns associated the choice of $h(z)$ and $\Omega_0$, we require some criteria to select among them. For convenience, we set $\Omega_0 = 1$.

To set $h(z)$, note that the local, areal growth, $\Omega$, obtained in any experiment must be bounded between $\Omega_{min} \le \Omega \le \Omega_{max}$.For $K=0$ surfaces, such as those we are considering, the cores of each singularity can never be accommodated in a finite range of growth and, so, those cores must be excised. Since these cores represent the smallest and largest swelling in any growth pattern associated with Eq. (\ref{eq:Omega}), we can formulate our search for an optimal growth pattern to be one that minimizes the area that must be excised around the singularities.

\begin{figure}
\includegraphics[width=3in]{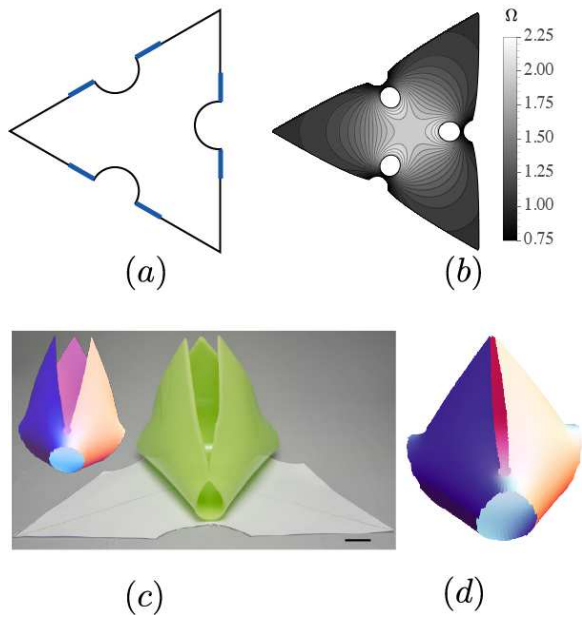}
\caption{\label{fig:tetra}  (color online) (a) Fold pattern for a tetrahedron with singularities removed. Identified regions of the sides are blue.  The corresponding growth pattern and domain with $B/R = D/R = 1.75$. The holes and boundary are chosen so that, $\Omega_{max}/\Omega_{min} = 3$. (c) Results of numerical minimization of (b) for a tetrahedron of thickness $0.04 R$ (inset) and result of folding (a) in poly(vinyl siloxane) with thickness $\approx 0.03 R$. (scale bar: 10 mm) (d) On the other hand, numerical minimization in which the each of the three positive singularities have Gaussian curvature $1.3 \pi$ closes imperfectly.}
\end{figure}

If we naively set $h(z)=0$, $\Omega(z,\bar{z})$ decreases to zero as $|z|$ becomes large. This immediately suggests the use of virtual singularities of opposite charge just outside the boundary of the domain, so that $\Omega(z,\bar{z}) \rightarrow 1$ as $|z| \rightarrow \infty$. The closer we are able to place these virtual singularities, the more uniform the growth will be. We proceed with the ansatz
\begin{equation}
|e^{h(z)}|^2 = |z^3-D^3|^{-1},
\end{equation}
where $D>R$ is chosen outside of the material domain. The closer $D$ is to $R$, the more uniform $\Omega(z,\overline{z})$ will be away from the vicinity of the singularities. The associated mapping is
\begin{equation}\label{eq:tetmap}
\partial g(z) = \left(\frac{z^3-D^3}{z^3-R^3}\right)^{1/2}.
\end{equation}
which corresponds to a wedge of angle $\pi$ near each singularity, consistent with identifying the two halves of each side of the triangle about its midpoint. With the virtual singularities, however, the entire side is not identified by the mapping. Instead, a length
\begin{equation}\label{eq:L}
L = R \int_1^{B/R} dx~\frac{|x^3-(D/R)^3|}{|x^3-1|}
\end{equation}
about the mid-point of each side, where $R < B \le D$ and $B$ is the distance to the boundary before growth along lines through the center (Fig. \ref{fig:tetra}a). This failure of identification has the propensity to alter the resulting shape from a completely closed tetrahedron (Fig. \ref{fig:tetra}c). Thus, Eq. (\ref{eq:L}) identifies one essential trade-off: we can make the growth pattern of the tetrahedron arbitrarily uniform by taking $D$ and $B \rightarrow R$ while simultaneously keeping the image singularities outside the boundary. However, we do this at the expense of shortening $L$. 

Beyond this, there is a second trade-off related to the resolution necessary to encode the growth pattern. The maximum growth occurs on the boundary of the positive Gaussian curvature singularities and the minimum near the virtual singularities. Therefore, the closer a virtual singularity is to a real singularity, the more rapidly the growth pattern changes between them. Mathematically, we can compute
\begin{equation}
\frac{dx}{d\Omega} = \frac{D^3-R^3}{3 [(\Omega-1)^4 (D^3-\Omega R^3)^2]^{1/3}}.
\end{equation}
How to put a bound on $dx/d\Omega$ clearly depends on details of a material system, but it is clear that the closer $D$ is to $R$, the more rapidly the growth pattern changes and, therefore, the more resolution necessary to describe the shape.

To corroborate our theoretical results, we have simulated the growth pattern in Eq. (\ref{eq:tetmap}) with $D/R = 1.75$ and boundary $B=D$ (Fig. \ref{fig:tetra}c, inset). The tetrahedron does not close. To corroborate this shape, we also folded a tetrahedron with thin elastic sheet of poly(vinyl siloxane) (Elite Double 32, Zhermack) using the pattern in Fig. \ref{fig:tetra}a. Attaching the relevant corners with narrow strips of silicone adhesive (ARclad® IS-8026, Adhesives Research, Inc.) results in a remarkably similar open tetrahedron (Fig. \ref{fig:tetra}c). One way to close the tetrahedron would be to use a larger range of growth, which would allow us to develop shapes equivalent to identifying a larger length along the boundaries of Fig. \ref{fig:tetra}a. Alternatively, increasing the Gaussian curvature at each of the three singularities to $\approx 1.3 \pi$ does result in an imperfectly closed tetrahedron (Fig. \ref{fig:tetra}d); this is equivalent to allowing the faces in Fig. \ref{fig:tetra}c to overlap slightly.

Even though our analysis only applies, strictly speaking, to surfaces with zero Gaussian curvature almost everywhere, we still believe that the results lend some insight into the optimal design of more general shapes by isotropic growth. In particular, one could imagine approximating a smooth surface using only singularities of Gaussian curvature, much as smooth charge densities approximate discrete charges in electrodynamics. Moreover, our notion of using virtual singularities corresponds with Chebyshev's principle, which states that optimal growth patterns (those having the smallest variation of $\ln \Omega$) also have constant $\Omega$ on their boundary \cite{grave}. This result can also be understood in terms of an electrostatics analogy: $\ln \Omega$ is analogous to the electric potential, and the boundaries act as conductors. Thus the charge density on the boundary adjusts to the presence of the Gaussian curvature within to minimize the total ``electric field.''

In summary, we have demonstrated a mapping between the buckling of developable surfaces due to nonuniform growth and the buckling of elastic sheets with a prescribed configuration of Volterra defects. This mapping provides new insights into the trade-offs between the growth pattern and the resulting shape: though there are many potential growth patterns corresponding to the same Gaussian curvature, and indeed ones with a very small range of growth are possible, the ones with largest range of growth will produce better approximations to the desired shape.

We acknowledge discussions with T. Hull, J. Silverberg, I. Cohen, R. Kamien and D. Sussman and funding from NSF DMR-0846582 (CDS) and ARO W911NF-11-1-0080 (NB and RCH).


\begin{thebibliography}{99}

\bibitem{sharon} Y. Klein, E. Efrati and E. Sharon, Science \textbf{315}, 1116 (2007).

\bibitem{hayward} J. Kim, J.A. Hanna, M. Byun, C.D. Santangelo and R.C. Hayward, Science \textbf{335}, 1201 (2012).

\bibitem{lewicka} M. Lewicka, L. Mahadevan, R. Pakzad, Proc. R. Soc. A \textbf{467}, 402 (2010).

\bibitem{mmm} M.M. M\"uller, M. Ben Amar, and J. Guven, Phys. Rev. Lett. \textbf{101}, 156104 (2008).

\bibitem{dipole} J. Guven, J. A. Hanna, O. Kahraman, M.M. M\"uller, Eur. Phys. J. E \textbf{36}, 106 (2013).

\bibitem{seung} H.S. Seung and D.R. Nelson, Phys. Rev. A \textbf{38}, 1005 (1988).

\bibitem{kupf} R. Kupferman, M. Moshe and J. Solomon, arXiv:1306.1624 (2013).

\bibitem{virus} T.T. Nguyen, R.F. Bruinsma, W.M. Gelbart, Phys. Rev. E \textbf{72}, 051923 (2005).

\bibitem{fullerene} T.A. Witten, H. Li, Europhys. Lett. \textbf{23}, 51 (1993).

\bibitem{clathrin} R.J. Mashl, R.F. Bruinsma, Biophys. J. \textbf{74}, 2862 (1998).

\bibitem{graphene} I.A. Ovid'ko, Rev. Adv. Mat. Sci. \textbf{32}, 1 (2012).

\bibitem{nelson} E.H. Yong, D.R. Nelson, L. Mahadevan, Phys. Rev. Lett. \textbf{111}, 177801 (2013).

\bibitem{volterra} V. Volterra, Ann. Scient. Ec. Norm. Sup. (Paris) \textbf{24}, 401 (1907); for a more modern view, see M. Kleman and J. Friedel, Rev. Mod. Phys. \textbf{80}, 61 (2008).

\bibitem{liang} Compare, for example, Eq. 4.8(a-b) of Ref. \cite{seung} and Eq. (1-2) of Ref. Liang, L. Mahadevan, Proc. Nat. Sci. Am. \textbf{106}, 22049 (2009).

\bibitem{marder} M. Marder and N. Papanicolaou, J. Stat. Phys. \textbf{125}, 1069 (2006).

\bibitem{muller} J. Guven and M.M. M\"uller, J. Phys. A \textbf{41}, 055203 (2008).

\bibitem{efrati} E. Efrati, E. Sharon and R. Kupferman, J. Mech. Phys. Sol. \textbf{57}, 762 (2009).

\bibitem{dias} M.A. Dias, J.A. Hanna and C.D. Santangelo, Phys. Rev. E \textbf{84}, 036603 (2011).

\bibitem{chen} B.G. Chen and C.D. Santangelo, Phys. Rev. E \textbf{82}, 056601 (2010).

\bibitem{grave} D.A. Grave, Crelle J. \textbf{140}, 247 (1911).




\end{thebibliography}
\end{document}